# Citation concept analysis (CCA) – A new form of citation analysis revealing the usefulness of concepts for other researchers illustrated by two exemplary case studies including classic books by Thomas S. Kuhn and Karl R. Popper

Lutz Bornmann (corresponding author)*, K. Brad Wray** & Robin Haunschild***

*Division for Science and Innovation Studies

Administrative Headquarters of the Max Planck Society

Hofgartenstr. 8,

80539 Munich, Germany.

Email: bornmann@gv.mpg.de

**Centre for Science Studies

Aarhus University

Ny Munkegade 118,

DK8000, Aarhus C, Denmark.

Email: kbwray@css.au.dk

*** Max Planck Institute for Solid State Research

Heisenbergstraße 1,

70569 Stuttgart, Germany.

Email: r.haunschild@fkf.mpg.de

**Abstract**

In recent years, the full text of papers are increasingly available electronically which opens up the possibility of quantitatively investigating citation contexts in more detail. In this study, we introduce a new form of citation analysis, which we call citation concept analysis (CCA). CCA is intended to reveal the cognitive impact certain concepts – published in a document – have on the citing authors. It counts the number of times the concepts are mentioned (cited) in the citation context of citing publications. We demonstrate the method using three classical examples: (1) *The structure of scientific revolutions* by Thomas S. Kuhn, (2) *The logic of scientific discovery – Logik der Forschung: Zur Erkenntnistheorie der modernen Naturwissenschaft* in German –, and (3) *Conjectures and refutations: the growth of scientific knowledge* by Karl R. Popper. It is not surprising – as our results show – that Kuhn's "paradigm" concept has had a significant impact. What is surprising is that it has had such a disproportionately larger impact than Kuhn's other concepts, e.g., "scientific revolution". The paradigm concept accounts for over 80% of the concept-related citations to Kuhn's work, and its impact is resilient across all disciplines and over time. With respect to Popper, "falsification" is the most used concept derived from his books. Falsification, after all, is the cornerstone of Popper's critical rationalism.

# 1 Introduction

The usual citation analyses focus on counting the number of citations to a focal paper (FP). To assess the usefulness of the FP, its citation count is compared with the citation count for a similar paper (SP, which has been published in the same research field and year). If the FP received significantly more citations than the SP, its impact is noteworthy: the FP seems to be more useful for other researchers than the SP. However, this simple counting and comparing of citations might lead to misleading results: Suppose the content of the SP has been extensively discussed in eight citing papers, with several mentions in different sections of each citing paper. In contrast, the FP has been mentioned only once in the introduction of 12 citing papers; the citations of the FP are always included in a long list of other cited publications. Although the FP received more citations than the SP, the SP seems to be the more useful publication for other researchers. The SP appears to be so important that other researchers deal with the content of the paper intensively.

Another example demonstrating possible deficits of simple citation counting and comparing concerns the meaning certain content in the publications has for other researchers. Publications may introduce important ideas, concepts, methods etc. and simple citation counting is not able to show why the publications are useful for researchers: do researchers focus on concept A, but ignore concept B from book C? Do researchers heavily use method D from book E, but have no use for method F?

Recently, Bu, Waltman, and Huang (2019) introduced the notion of a multi-dimensional perspective on citation impact which goes beyond simple citation counts, the usual one-dimensional perspective. The authors argue for considering other aspects, which can be derived from citation data, in impact analyses for assessing different kinds of research impact. They introduce two indicators, depth and dependence, based on co-citations and bibliographic coupling which reveal how innovative certain publications are in a narrow research area or cross-disciplinary application. However, the multi-dimensional perspective on citation impact is not new. A decade ago, Bornmann and Daniel (2008b) published an extensive overview of studies investigating citation context and citation content of cited and citing publications as well as motivations of citing authors (see also Case & Higgins, 2000; Cronin, 1984; Ding et al., 2014; Liu, 1993; Small, 1982). These studies revealed many different reasons for citing publications and developed schemes for categorizing cited publications with regard to their importance for the citing author. Based on these studies the only conclusion which can be drawn is that the process of citing and the meaning of citations are multi-dimensional phenomena (Tahamtan & Bornmann, 2018). The acceptance of the multi-dimensional perspective on citation impact might be contrasted with the use of citations in research evaluation processes, since it is based on the notion, rooted in the normative view on citing publications (Merton, 1957), that citations reflect cognitive influence.

Various possible ways are conceivable to improve citation analyses in the direction of more detailed meaning. For example, ideas, concepts, methods etc. from certain publications may have been labeled as "discoveries" by citing researchers (Small, Tseng, & Patek, 2017) which is not visible by simple citation counting. An obvious way of improving citation impact measurements alongside the normative view is to count citations of publications with indications of the cognitive influence. In this study, we therefore introduce a new form of citation analysis, which we call citation concept analysis (CCA). CCA is intended to reveal the cognitive impact certain concepts have on the citing authors. It counts the number of times the concepts are mentioned (cited) in the citation context of citing publications. Of course, CCA only makes sense with publications introducing at least one important concept, which can be detected reliably in the citation context of citing publications.

In this study, we explain and demonstrate the use of CCA by using different editions of several publications introducing several concepts each: *The structure of scientific revolutions* by Kuhn (1962), *The logic of scientific discovery* (Popper, 1959) – *Logik der*

*Forschung: Zur Erkenntnistheorie der modernen Naturwissenschaft* (Popper, 1934) in German – and *Conjectures and refutation: the growth of scientific knowledge* by Popper (1962).[1] The books explain several concepts each, e.g., "scientific revolution" or "falsification", which can be searched in the citation context of citing publications. We are interested in how useful the concepts from the books are for the citing authors. Since the books represent two contrary understandings of scientific progress, CCA results might also reveal how scientific progress is differently interpreted in various disciplines and time periods.

In a second part of the empirical analysis, following the proposal by Small, Boyack, and Klavans (2019), we searched for hedging word strings expressing uncertainty (e.g., "may" and "could") in citation context data of Kuhn (1962) and Popper (1934, 1959, 1962). We are interested in which concepts are more frequently associated than other concepts with uncertainty: which concepts are interpreted as scientifically proven and which not?

## 2 Theoretical reflections on citation concept analysis (CCA)

Since the use of citations in measuring research gained a certain level of attention, researchers have started to develop theories of the citation process. According to Sandström (2014), these theories try "to explain why author *x* cite[s] article *a* at time *t*? What factors should be considered when we discuss why researchers cite back to former literature?" (p. 59). Tahamtan and Bornmann (2018) published a conceptual overview of the literature dealing with the process of citing documents. They identified three core elements in this process: the context of the cited document, processes from selection to citation of documents, and the context of the citing document. Tahamtan and Bornmann (2018) not only considered empirical studies on the citing process, but also theories of the citing process which have been proposed hitherto. One of these theories is of special importance for the CCA (see below).

---

[1] In this study, we considered not only the first editions of the books, but also later editions.

Robert K. Merton proposed the first theory of citing: the normative theory. Merton (1973) discusses citations in the context of various norms of science, which are part of the ethos of science. One of these norms "communism" predetermines that researchers credit the original research of other researchers, and citations are a formal process in publishing research to do that. Since citations point to published research which has been highlighted by publishing researchers in their reference lists, Merton (1973) sees citations as a part of the reward system of science, which is one of the reasons why citations can be used for research evaluation purposes. The second theory of citing is in sharp contrast to the normative theory. Social constructivists (e.g., Gilbert, 1977) claim that researchers cite publications for rhetorical and strategic reasons whereby these reasons do not, or scarcely, depend on the quality of the cited publication (Case & Higgins, 2000). For example, in view of the social constructivist theory of citing, researchers cite strategically certain publications which support their own claims, (and avoid to cite publications with contrary results; Teplitskiy, Duede, Menietti, & Lakhani, 2018). Social constructivists doubt that citation decisions can be explained with only one dominant reason, as proposed by the normative theory of citing. Several reasons exist – also for single citation decisions – which rarely has something to do with the quality of the cited research.

Small (1978) proposed a third view of citing which can be connected to both the normative and social constructivist view. Small (1978) interprets citations as an occurrence of symbol usage. In the analysis of highly cited documents in chemistry, Small (1978) found "a high degree of uniformity … in the association of specific concepts with specific documents" (p. 327) and concluded that certain documents (important or highly-cited documents) may be seen as "'standard symbols' for particular ideas, methods, and experimental data" (p. 327). These concept symbols can be used by the citing authors to credit original research, following the normative theory of citing, and for rhetorical and strategic reasons, following the constructivist theory of citing. For example, Kuhn (1962) can be cited because the citing author uses the paradigm concept in his or her own research. Another reason – in the spirit of the constructivist theory – would be to cite Kuhn (1962) because the citing author believes that the reference to Kuhn (1962) has positive effects in the later assessment of his or her manuscript, e.g., because possible reviewers or chief editors are proponents of concepts introduced by Kuhn (1962).

The theoretical roots of the CCA are located in the third view on the citing process proposed by Small (1978). In the first step of the CCA, certain concepts (concept symbols) are identified which are connected with a certain (landmark) publication. In the second step, search terms are defined which can be used to identify the concepts in so-called citances. A citance is defined as a "single sentence in which the citation was made" (Small, 2018, p. 463). In the third step, the statistical analysis, the search terms are used to determine the number of concept occurrences in citances. Small (1978) already proposed to investigate the text around citations to identify the particular idea which connects the citing authors with the cited document. The availability of large amounts of citation context data in bibliometric databases such as Microsoft Academic (MA) makes it possible to investigate concept symbols in large datasets.

## 3  Importance of the books published by Thomas S. Kuhn and Karl R. Popper used as examples in this study

The importance of *The structure of scientific revolutions* by Kuhn, *The logic of scientific discovery – Logik der Forschung: Zur Erkenntnistheorie der modernen Naturwissenschaft* in German –, and *Conjectures and refutation: the growth of scientific knowledge* by Popper cannot be overstated. Popper's *Logic of scientific discovery* marked the beginning of his academic career. The book was completed when he was working as a high school teacher in Austria in the 1930s. But it played an integral role in securing his first position at a university in New Zealand. In the book, Popper (1959) presents the central or organising idea of his whole career, falsificationism. He argues that, contrary to the reigning orthodoxy, scientists cannot confirm hypotheses or theories by testing them. Rather, from a logical point of view, when they test hypotheses and theories by deriving predictions from them, all they can do is attempt to falsify them. If a theory is in fact falsified, if it entails a false prediction, then it should be discarded or modified. If it is not falsified, if the predictions from it are corroborated, then it can continue to be accepted. But the acceptance of a hypothesis or theory, Popper insists, is always tentative. Some future test may prove the theory false. One further important reason why Popper's book is regarded as significant is the role it played in challenging the then-orthodox view, *Logical Positivism*. The *Positivists* were based principally in Vienna, and it was in that milieu that Popper developed his philosophy of science. Popper would later claim that it was he who killed positivism (Popper, 1992, § 17).

All the other concepts we examine from Popper's books are intimately connected to falsificationism: verification, induction, demarcation, and probability. The positivists were seeking to develop a logic of verification. Popper, though, argued that it is not possible to verify a hypothesis or theory. Popper argued that when one infers the truth of a theory from an observation predicted from theory, one is committing a logical fallacy, the fallacy of affirming the consequent. The positivists also aimed to develop a logic of induction. Popper, though, argued that there is no logic of induction. The only way that scientists can prove a claim is by deduction. Specifically, when a scientist derives a prediction from a theory, and the world is not as the theory predicts, then the scientist can deductively infer that the theory is false. This is an instance of the inference pattern *modus tollens*. Demarcation is another important concept for Popper. He believes that falsifiability is the feature that distinguishes scientific theories, like Isaac Newton's and Albert Einstein's, from pseudo-scientific theories, like Sigmund Freud's theory of psychology or Karl Marx's theory of history. Falsifiability is thus Popper's demarcation principle. Finally, probability is another concept central to Popper's thinking. His studies of probability were motivated by his concern to develop an understanding of corroboration. Even though vindicated predictions could not prove a theory is true, they did corroborate it, that is, offer some sort of support for it. Popper's studies of probabilities were motivated, in part, by a concern to bring rigour to the notion of corroboration.

Kuhn's (1962) *Structure of Scientific Revolutions* is also undeniably an important contribution to the philosophy of science. It was responsible for two of the citation peaks in the philosophy of science from 1900 to 1970 (see Wray & Bornmann, 2015). And its effects were wide ranging (see Wray, 2017). Kuhn (1962) played a crucial role in developing the successor project in the philosophy of science to the then-dominant view, *Logical Positivism*. Kuhn (1962) argued that the development of scientific fields was not a steady progression, but involved the periodic interruption of radical changes of theory, what he called paradigm changes. These changes involved significant changes to the basic assumptions about the nature of the world. Kuhn (1962) thus developed a cyclical account of the growth of science. On the one hand, his book was quite quickly greeted with enthusiasm as it seemed to offer insight into how science really works, drawing on a study of the history of science. On the other hand, it seemed to undermine the presumed rationality of science. Kuhn (1962) compared paradigm changes to religious conversions and gestalt shifts, processes that are typically regarded as non-rational. He also spoke of scientific fields as in crises. And competing theories were described as incommensurable. These were hardly comforting words to scientists or philosophers of science. But they certainly created excitement across the academic world. Even in literary theory students were encouraged to read *The structure of scientific revolutions* by Kuhn (1962).

The other concepts we examine in Kuhn's *Structure* are related to the notion of paradigm change. They include normal science, theory change, and anomaly. Normal science is the phase of scientific research that falls between paradigm changes. In fact, normal science is the sort of research that most scientists engage in most of their careers. It involves the uncritical application or extension of an accepted paradigm or theory. And theory change is just an alternative term used to describe paradigm changes. Periods of normal science are punctuated by changes of theory. And anomalies are the problematic observations that scientists are unable to account for given the resources of the accepted theory. They are thus the cause of revolutionary changes of theory.

## 4 Literature review (of citation context and content studies)

Since the citation process is scarcely standardized (Cronin, 1982), the adoption of a multi-dimensional perspective on citations seems reasonable. As outlined by Tahamtan and Bornmann (2018), several reasons and decision rules exist for citing why it is not clear what is being measured with citations: cognitive influence, citation circles, Matthew effects, field-specific impact patterns, rhetorical persuasion etc. The results of the empirical study by Judge, Cable, Colbert, and Rynes (2007) based on 21 management journals reveals that "universalistic, particularistic, and mixed universalistic-particularistic characteristics all play significant roles in the extent to which research articles in the field of management are cited" (p. 500). Already at the end of the 1980s based on various citation studies Cano (1989) concluded that "neither citation behavior, nor citation use or citation impact has been satisfactorily explained" (p. 284). We think that the situation has not changed substantially since then, despite considerable research efforts in this area.

A decade ago, Bornmann and Daniel (2008b) reported the results of the previous empirical literature dealing with the multi-dimensional perspective on citation impact. The previous studies can be assigned to three groups: (1) Citation context studies are based on a text analysis of one or two sentences around an in-text citation and aim to classify the citation (e.g., as "confirmative" or "negational"). According to Boyack, van Eck, Colavizza, and Waltman (2018) an in-text citation "is a mention of a reference within the full text of a document. A reference can be mentioned one or more times in a document. Each mention is an in-text citation" (p. 60). (2) Citation content analyses investigate the semantic content of the citing papers to characterize the cited publication. (3) Citer motivation interviews or surveys identify citer motives by surveying or interviewing citing authors. Following this classification, the empirical parts of the current study are citation context analyses since the text around the in-text citation is searched for the existence of certain search terms or concepts.

In recent years, the situation changed for undertaking citation context or content analyses. In the past, it was very time-consuming and resource intensive to undertake these studies. The analyses had to be done manually by searching for certain in-text citations and categorizing their content. A good example of these comprehenisve studies is the citation context study undertaken by Bornmann and Daniel (2008a). However, the situation changed with the growing dissemination of large datasets as outlined by Bertin, Atanassova, Sugimoto, and Lariviere (2016): "large-scale datasets containing the full-text of scholarly documents have become available and initiatives have been undertaken to provide appropriate mark-up for citation context analysis" (p. 1418). Bornmann, Haunschild, and Hug (2018) compared keywords of citing papers and keywords of cited papers with terms in the citation context of the citing papers. Due to the increasing availability of datasets, an increasing number of studies have been published using, e.g., datasets from PubMed Central, Public Library of Science (PLOS), MA, and Elsevier.

One popular way of "adding depth and value to the citation metric" (Carroll, 2016, p. 1329) is to determine the frequency with which a certain paper is cited within single citing publications. In a recent study, Zhao and Strotmann (2018) investigated the frequency of these in-text citations and the sections in which publications have been cited. They investigated the rank position changes of single authors by comparing usual citation counting with in-text frequency-weighted citation counting and including or excluding citations from certain sections. Zhao and Strotmann (2018) found that essential citation impact can be determined especially when citations from introductory and background sections are removed. However, the best approach to identify essential citations is the combination of frequency-weighted citation counting with citation counting in certain sections of the citing papers.

Pak, Yu, and Wang (2018) also focused on the frequency of in-text citations in scientific papers. They investigated about 5000 papers, with the document type "article", published in five journals from computer science, biology, physics, and chemistry. The authors propose that the results can be interpreted as an instance of the 80/20 rule in citation data: about 20% of the cited references might be more important, since multiply mentioned, for the citing authors than the remaining 80% of the cited references. Using citations in papers published in the *Journal of Informetrics* as their dataset, Hu, Chen, and Liu (2015) analyzed the reasons for multiple citations of single publications: "By comparing the contexts of two kinds of citations, the first-time citations and the succeeding citations, we found that, for a specific reference, its first-time citation is usually not as intentional as the succeeding citations. Just because of the relative importance of the succeeding citations compared to the first-time citation, recurring citations are reasonable and necessary" (p. 221).

The study by Bertin, Atanassova, Gingras, and Larivière (2015) focusses on the distribution of cited references across the sections in papers. The study is based on the full text of papers published in PLOS journals. The results show that the introduction and discussion sections account for most of the cited references. The introduction frames the research question of the study, and the discussion puts the results of the study in the perspective of previous studies. Bertin et al. (2015) also found that "the age of cited references varies by section, with older references being found in the methods and more recent references in the discussion" (p. 164). This result is in accordance with the finding by van Noorden, Maher, and Nuzzo (2014) that many highly cited papers are papers proposing research methods – methods seem to be used conservatively and thus for a long time. The study by Bertin et al. (2015) also reveals that disagreement with previous research is less common than agreement; agreement is indicated mostly at the end of a paper.

Doslu and Bingol (2016) applied the citation context approach to identify fundamental and important papers in a context-sensitive way, by ranking important papers in a topic-specific way. The study by Small et al. (2017) used a similar approach as we do in this current study. They searched in citing sentences for the truncated string "*discover*" to determine whether the cited paper is a revolutionary work leading to a discovery. However, the results of the study point out that this approach to citation context analysis cannot be used for reliably finding discoveries: "While we have found that the appearance of 'discovery words' is not a reliable indicator of whether a given citance [which is the sentence including a specific citation] describes a discovery (the success rate is about 46%), it is relatively easy to differentiate these manually by inspecting a sample of citances. Because the majority of false hits are methodologies and tools for making discoveries, rather than actual scientific discoveries, the vocabulary used in citances in such cases has been shown to be highly instrumental, and to differ markedly from the vocabulary used when actual discoveries are cited" (p. 56).

Boyack et al. (2018) deal with field-specific differences in in-text citations in many millions of full texts of papers (from PubMed Central). They found "large field-level differences that are reflected in [the] position within the text, citation interval (or reference age), and citation counts of references" (p. 59). The results further show that publications cited only once in citing publications tend to be more highly cited than publications cited multiple times. The latter result might be in disagreement with the results of many previous studies. Jha, Jbara, Qazvinian, and Radev (2016) used as their dataset the ACL Anthology Network (AAN), which provides the full text of papers published in the field of Natural Language Processing (NLP). The authors investigated the purpose and polarity of citations. They found that the purpose categories were: 14.7% criticism, 8.5% comparison, 17.7% use, 7% substantiation, 5% basis, and 47% other. 30% had a positive polarity, 12% a negative one, and 58% a neutral one.

Three studies have used various approaches for studying the use of content from cited papers more deeply. Greenberg (2009) investigated how a belief system evolves in a scientific community, "the belief that β amyloid, a protein accumulated in the brain in Alzheimer's disease, is produced by and injures skeletal muscle of patients with inclusion body myositis" (Greenberg, 2009). Greenberg (2009) searched for statements addressing this belief in papers indexed in PubMed related to the β amyloid. Citation networks have been analyzed by social network and graph theory: "the network contained 242 papers and 675 citations addressing the belief, with 220553 citation paths supporting it. Unfounded authority was established by citation bias against papers that refuted or weakened the belief; amplification, the marked expansion of the belief system by papers presenting no data addressing it; and forms of invention such as the conversion of hypothesis into fact through citation alone" (Greenberg, 2009).

Hammarfelt (2011) investigated the impact of a highly cited publication in humanities, Walter Benjamin's *Illuminations* (Benjamin, 2007). Hammarfelt (2011) used a new method for measuring detailed impact by additionally considering the sections and cited pages. The results show that "a majority of the citations refer to a few specific essays in the collection 'The Work of Art in the Age of Mechanical Reproduction,' 'The Storyteller,' and 'Theses on the Philosophy of History.' The single most cited page, with 327 citations, is page 217, which is the first page of 'The Work of Art in the Age of Mechanical Reproduction'" (p. 826). Solomona, Youtieb, Carleyc, and Porter (submitted for review) explored the citations the seminal US National Academies consensus report, *How People Learn* (HPL), received (National Research Council, 2000). When they analyzed the actual content in HPL to which those citing publications referred, they "found them to be more disciplinary than might have been supposed. The Education publications that cited HPL were overwhelmingly more likely to cite content derived from the Education literature rather than from CogSci, whereas the CogSci publications that cited HPL were more likely to cite content derived from the CogSci literature". Sieweke (2014) used a similar approach as we did in the current study to investigate the influence the French sociologist Pierre Bourdieu has in the field of management and organization research. The author did not only analyze the citations Bourdieu's publications received over time, but also which contents from the publications were of greatest interest. The results show that "the three concepts of capital, habitus and field combined cover almost half of the citations" (Sieweke, 2014, p. 537).

# 5 Methods

## 5.1 Datasets

In this study, we used citation context data from MA (Sinha et al., 2015), see also https://aka.ms/msracad. We searched for the book titles to find all database entries which belong to the different editions of the three books by Kuhn and Popper. For Kuhn, the editions from 1962, 1970, and 1996 have two different entries, each, while the 50[th] anniversary edition from 2012 has only one entry. For Popper, we found one database entry each for the German and English editions of *Logik der Forschung: Zur Erkenntnistheorie der modernen Naturwissenschaft* and *The logic of scientific discovery*, respectively, and we found a single entry for *Conjectures and refutations: the growth of scientific knowledge*.

Certain terms – key concepts from Kuhn and Popper – have been searched in citances. We use the expression "citance" in this study although the citation context in MA frequently includes more than one sentence, or only a part of a sentence. The key concepts that we examine in this study from Kuhn's book are the core concepts of his cyclical model of scientific change: paradigm, normal science, crisis, incommensurability, scientific revolutions, and structure. The key concepts from Popper's books that we examine are derived from his two books: they are all related to the chapter titles. These are concepts that figured importantly in a number of debates in the philosophy of science: induction, falsification, demarcation, corroboration, and probability.

We searched the concepts in the citances in the following manner: In the first step, punctuation characters (",", ".", ";", and ":") were removed from the citances and substituted with whitespaces, and all citances were converted to lower case characters. Second, a word frequency analysis was performed on the citances. Next, frequently occurring words, which describe common concepts employed by Kuhn (1962) or Popper (1934, 1959, 1962) were used to construct the search terms. For some concepts, the occurrence of the search terms was too low, and those concepts (e.g., "theory change" for Kuhn and "verifiability" for Popper) were dropped. We obtained a set of seven concepts for Kuhn and five concepts for Popper with ample amount of hits in the citances. The concepts and the corresponding search terms for Kuhn and Popper are shown in Table 1. In the case of the concepts "scientific revolution" and "structure", the citances are excluded where the book title itself occurs. In the case of Popper, German search terms were employed, too.

Table 1. Concepts and corresponding search terms for Kuhn and Popper

| Concept | Search term(s) |
|---|---|
| Kuhn | |
| Scientific revolution | "scientific revolution*" |
| Paradigm | "paradigm*" |
| Normal science | "normal science" |
| Structure | "structure" |
| Anomaly | "anomalies", "anomaly", and "anomaliety" |
| Incommensurability | "incommensurability" and "incommensurable" |
| Crisis | "crisis" and "crises" |
| Popper | |
| Induction | "induction" and "induktion" |
| Falsification | "falsif*" |
| Demarcation | "demarcation" and "abgrenzung" |
| Corroboration | "corroborate", "bestätigen", "bestatigen", "bestaetigen", "corroboration", "bestätigung", "bestatigung", and "bestaetigung" |
| Probability | "probability" and "wahrscheinlichkeit" |

Note. The asterisk is a truncation symbol.

Unfortunately, the citation context is not available for all citing papers in MA. For Kuhn's book, we found 38,474 citing papers, 4,710 of them have citation context information available, and 4,688 of the citing papers with citation context were assigned to at least one level 0 field of study (FOS) – the highest level in the field categorization scheme used by MA. In total, we found 2,441 papers citing Kuhn's book which contained at least one of the search terms in Table 1. For Popper's books, we found 14,829 citing papers, 2,129 of them have citation context information available, and 2,122 of the citing papers with citation context were assigned to at least one level 0 FOS. In total, we found 663 papers citing Popper's books which contained at least one of the search terms in Table 1. Some of the citing papers referenced Popper and/or Kuhn multiple times and may be assigned to more than a single level 0 FOS.

In this study, an approach proposed by Small (2018) and Small et al. (2019) is used to measure uncertainty associated with concepts. Small et al. (2019) interpret hedging as an expression of uncertainty which is the inverse of certainty. The authors propose to use certain hedging words to measure uncertainty associated with cited publications. Small (2018) reports that the hedging word "may" can be more frequently observed in citation contexts for lower cited than for more highly-cited papers. Small et al. (2019) used the open access subset from PubMed Central and searched in the provided citation context data for hedging word strings (e.g., "may" and "could"). The authors note that "hedging does not assert that the paper is wrong, but only suggests that uncertainty surrounds some aspect of the ideas put forward" (Small et al., 2019). They computed hedging rates "separately for method and non-method papers, the latter being more frequently hedged. Rates of hedging are found to be higher for papers with fewer citances, suggesting that the certainty of scientific results is directly related to citation frequency" (Small et al., 2019). According to the results by Chen, Song, and Heo (2018), who published a conceptual framework for the study of uncertainties in publications, fields with the highest rates of uncertainty are psychology, business, management, accounting, social sciences, economics, econometrics, and finance. Low rates can be found in mathematics, material science, and chemistry. Hyland (1996) distinguishes between reader-motivated (e.g., "believe", "suggest", and "analogy") and content-motivated (e.g., "generally", "almost", "might", and "probable") hedging words. A thorough discussion of problems in undertaking uncertainty analyses can be found in Small et al. (2019).

Henry Small – one of the co-authors of Small et al. (2019) – provided us with his initial hedging words list for measuring uncertainty in his study. The list includes the following terms: not clear, no clear, appears, possibility, seems, speculated, to some extent, impression, sometimes, perhaps, not known, seem, apparently, tends, not necessarily, preliminary, contingent, could, doubt, explore, feel, hope, hopeful, hopefully, likely, may, might, nevertheless, nonetheless, not known, opportunity, plausible, possible, possibly, potential, potentially, probable, probably, projected, promise, promising, questions, risky, speculative, suspect, uncertain, unclear, unknown, unsolved, whether, and yet to be determined. We checked in the citances of our data set how frequently the terms occur. For the statistical analyses, we selected only those terms, which occur very frequently: "may", "possible", "could", "questions", "might", "whether", "potential", "seems", "perhaps", "likely", and "sometimes". The resulting set was reduced when we eliminated terms for which we observed that they are frequently not used to express uncertainty in the citances. The final set is as follows: "like", "may", "could", "questions", "might", "potential", "seems", "perhaps", "likely", and "sometimes". The terms "possible" and "whether" were not used because many citances did not express uncertainty when these terms were used.

Besides measuring uncertainty, we also tried sentiment analyses of citation context data, but the results were not very promising. Thus, we decided not to include the results of the analysis in this manuscript. However, it would be interesting to explore whether certain concepts received a positive or negative connotation by the citing authors (similar to the uncertainty analysis).

**5.2  Statistics**

In this study, we basically count how often certain terms, i.e. concepts (e.g., "paradigm"), are mentioned in citances of citing publications. As more than one concept is frequently mentioned in single citances, it is necessary to consider in the statistical analyses the possibility that concepts are mentioned multiple times. The meta-data in MA allows us to analyze differences between fields and time periods in citing certain concepts by using the FOSs and publication years of the citing publications. The relationship of concepts and fields, or time periods, can be represented in a contingency table whereby the citations of concepts are dependent variables, and fields and time periods are independent variables. We used the Stata command *mrtab* (Jann, 2005) to analyze multiple mentions of concepts depending on FOS and publication year. On the one hand, we inspect differences in percentages of concept mentions between different FOSs and publication years as indications of effect size (Cumming & Calin-Jageman, 2016). On the other hand, we calculate statistical significance tests in two-way tables: (1) We perform overall chi-square tests to investigate the overall relationship between the cited concept and FOS, or publication year. (2) We perform a series of separate chi-square tests for each concept whereby the $p$ values are adjusted to account for simultaneous testing.

# 6   Results

We undertook a CCA for several landmark publications in philosophy of science: Kuhn (1962) and Popper (1934, 1959, 1962). First, we performed the CCA for various FOSs and time periods. Second, we analysed the (un-)certainty connected to the concepts. The results are presented in the following sub-sections.

## 6.1 First empirical part: citations of concepts in various fields of study (FOSs) and time periods

The results of the CCA for Kuhn (1962) are shown in Table 2. For each concept, the number of occurrences (citances) in each FOS are presented (absolute numbers and in percent). Since multiple mentions of different concepts in a single citance are possible, the total column percentages are greater than 100% for most of the FOS. The table reports column percentages: we assume that the concept mentions are dependent on the FOS.

Table 2. Citation concept analysis of *The structure of scientific revolutions* by Kuhn. How frequently have concepts (the concepts are decreasingly sorted by the column "Total") been cited in various fields of study (FOSs)?

| Concept | | Field of study (FOS) | | | | | | | | | | | | | | | | | | | |
|---|---|---|---|---|---|---|---|---|---|---|---|---|---|---|---|---|---|---|---|---|---|
| | | Art | Biology | Business | Chemistry | Computer science | Economics | Engineering | Environment | Geography | Geology | History | Material science | Mathematics | Medicine | Philosophy | Physics | Political science | Psychology | Sociology | Total |
| Paradigm | $n$ | 10 | 95 | 72 | 8 | 308 | 221 | 142 | 9 | 35 | 7 | 1 | 6 | 77 | 101 | 42 | 44 | 210 | 503 | 403 | 2294 |
| | % | 83.33 | 83.33 | 85.71 | 72.73 | 77.97 | 76.47 | 89.31 | 75 | 92.11 | 70 | 100 | 85.71 | 74.04 | 87.07 | 77.78 | 74.58 | 83.33 | 84.11 | 83.78 | 82.05 |
| Scientific revolution | $n$ | 1 | 18 | 4 | 2 | 70 | 35 | 14 | 4 | 4 | 1 | 0 | 2 | 17 | 12 | 6 | 14 | 32 | 75 | 48 | 359 |
| | % | 8.33 | 15.79 | 4.76 | 18.18 | 17.72 | 12.11 | 8.81 | 33.33 | 10.53 | 10 | 0 | 28.57 | 16.35 | 10.34 | 11.11 | 23.73 | 12.7 | 12.54 | 9.98 | 12.84 |
| Normal science | $n$ | 0 | 7 | 8 | 1 | 44 | 39 | 9 | 3 | 3 | 2 | 0 | 0 | 16 | 5 | 4 | 2 | 27 | 36 | 48 | 254 |
| | % | 0 | 6.14 | 9.52 | 9.09 | 11.14 | 13.49 | 5.66 | 25 | 7.89 | 20 | 0 | 0 | 15.38 | 4.31 | 7.41 | 3.39 | 10.71 | 6.02 | 9.98 | 9.08 |
| Structure | $n$ | 0 | 11 | 4 | 1 | 24 | 12 | 7 | 0 | 4 | 0 | 0 | 2 | 5 | 6 | 4 | 2 | 19 | 33 | 26 | 160 |
| | % | 0 | 9.65 | 4.76 | 9.09 | 6.08 | 4.15 | 4.4 | 0 | 10.53 | 0 | 0 | 28.57 | 4.81 | 5.17 | 7.41 | 3.39 | 7.54 | 5.52 | 5.41 | 5.72 |
| Anomaly | $n$ | 0 | 2 | 2 | 0 | 14 | 21 | 2 | 1 | 0 | 1 | 0 | 0 | 7 | 3 | 1 | 1 | 11 | 30 | 17 | 113 |
| | % | 0 | 1.75 | 2.38 | 0 | 3.54 | 7.27 | 1.26 | 8.33 | 0 | 10 | 0 | 0 | 6.73 | 2.59 | 1.85 | 1.69 | 4.37 | 5.02 | 3.53 | 4.04 |
| Incommensurability | $n$ | 0 | 1 | 1 | 1 | 14 | 9 | 4 | 0 | 1 | 0 | 0 | 0 | 9 | 2 | 4 | 4 | 6 | 18 | 25 | 99 |
| | % | 0 | 0.88 | 1.19 | 9.09 | 3.54 | 3.11 | 2.52 | 0 | 2.63 | 0 | 0 | 0 | 8.65 | 1.72 | 7.41 | 6.78 | 2.38 | 3.01 | 5.2 | 3.54 |
| Crisis | $n$ | 1 | 6 | 1 | 0 | 6 | 6 | 5 | 0 | 0 | 1 | 0 | 0 | 1 | 5 | 1 | 1 | 6 | 11 | 8 | 59 |
| | % | 8.33 | 5.26 | 1.19 | 0 | 1.52 | 2.08 | 3.14 | 0 | 0 | 10 | 0 | 0 | 0.96 | 4.31 | 1.85 | 1.69 | 2.38 | 1.84 | 1.66 | 2.11 |
| Total | $N$ | 12 | 140 | 92 | 13 | 480 | 343 | 183 | 17 | 47 | 12 | 1 | 10 | 132 | 134 | 62 | 68 | 311 | 706 | 575 | 3338 |
| | % | 100 | 122.81 | 109.52 | 118.18 | 121.52 | 118.69 | 115.09 | 141.67 | 123.68 | 120 | 100 | 142.86 | 126.92 | 115.52 | 114.81 | 115.25 | 123.41 | 118.06 | 119.54 | 119.38 |
| Cases | $N$ | 12 | 114 | 84 | 11 | 395 | 289 | 159 | 12 | 38 | 10 | 1 | 7 | 104 | 116 | 54 | 59 | 252 | 598 | 481 | 2796 |

Notes. Results of chi-square tests for single concepts (rows): paradigm: $\chi^2=34.50$, $p=0.076$; scientific revolution: $\chi^2=35.35$, $p=0.06$; normal science: $\chi^2=38.26$, $p=0.025$; structure: $\chi^2=18.87$, $p=1$; anomaly: $\chi^2=23.64$, $p=1$; incommensurability: $\chi^2=25.6$, $p=0.764$; crisis: $\chi^2=18.29$, $p=1$. Result of the overall chi-square test: $\chi^2=733.61$, $p=0.092$.

As the results in the table reveal, "paradigm" (with 82.05%) is the concept which is most frequently mentioned in citances of Kuhn (1962). Since this is the case in all FOSs, it seems that the book is mainly associated with the paradigm concept. With a large difference in numbers (around 70 percentage points), "paradigm" is followed by "scientific revolution" with 12.84%.

In the notes of Table 2, the results of various chi-square tests are shown. The results for single concepts reveal whether the relationship between concept mentions and FOS is statistically significant (in cases where $p <. 05$). As the results in the table show, there is only one statistically significant result, namely for the concept "normal science". For example, "normal science" has been more frequently mentioned in mathematics (with 15.38%) than in physics (with 3.39%). Below the table, there also is the result of the overall chi-square test which shows whether the relationship between concepts and FOS is statistically significant in general. As the test result points out, the general relationship is statistically not significant.

Table 3. Citation concept analysis of *The structure of scientific revolutions* by Kuhn: How frequently have concepts (the concepts are decreasingly sorted by the column "Total") been cited in various time periods (publication years)?

| Concept | | Time period (publication years) | | | | |
|---|---|---|---|---|---|---|
| | | <2000 | 2000-2005 | 2006-2010 | 2011-2018 | Total |
| Paradigm | $n$ | 391 | 422 | 771 | 710 | 2294 |
| | % | 82.14 | 79.47 | 82.55 | 83.04 | 82.05 |
| Scientific revolution | $n$ | 58 | 66 | 121 | 114 | 359 |
| | % | 12.18 | 12.43 | 12.96 | 13.33 | 12.84 |
| Normal science | $n$ | 37 | 46 | 103 | 68 | 254 |
| | % | 7.77 | 8.66 | 11.03 | 7.95 | 9.08 |
| Structure | $n$ | 23 | 34 | 54 | 49 | 160 |
| | % | 4.83 | 6.4 | 5.78 | 5.73 | 5.72 |
| Anomaly | $n$ | 24 | 27 | 30 | 32 | 113 |
| | % | 5.04 | 5.08 | 3.21 | 3.74 | 4.04 |
| Incommensurability | $n$ | 14 | 23 | 31 | 31 | 99 |
| | % | 2.94 | 4.33 | 3.32 | 3.63 | 3.54 |
| Crisis | $n$ | 6 | 12 | 18 | 23 | 59 |
| | % | 1.26 | 2.26 | 1.93 | 2.69 | 2.11 |
| Total | $N$ | 553 | 630 | 1128 | 1027 | 3338 |
| | % | 116.18 | 118.64 | 120.77 | 120.12 | 119.38 |
| Cases | $N$ | 476 | 531 | 934 | 855 | 2796 |

Notes. Results of chi-square tests for single concepts (rows): paradigm: $\chi^2=3.13$, $p=1$; scientific revolution: $\chi^2=0.46$, $p=1$; normal science: $\chi^2=6.7$, $p=0.574$; structure: $\chi^2=1.16$, $p=1$; anomaly: $\chi^2=4.57$, $p=1$, incommensurability: $\chi^2=1.63$, $p=1$; crisis: $\chi^2=3.26$, $p=1$. Result of the overall chi-square test: $\chi^2=120.47$, $p=0.321$.

Table 3 reveals the relationship between concepts and time periods for *The structure of scientific revolutions* (Kuhn, 1962). Have certain concepts been increasingly mentioned over time, but others not? As the results of the statistical significance tests show, no results are statistically significant. Thus, the use of the concepts in the citing literature scarcely changes over time.

Table 4. Citation concept analysis of *The logic of scientific discovery* (in German: *Logik der Forschung: Zur Erkenntnistheorie der modernen Naturwissenschaft*) and *Conjectures and refutations: the growth of scientific knowledge* by Popper. How frequently have concepts (the concepts are decreasingly sorted by the column "Total") been cited in various fields of study (FOSs)?

| Concept | | Field of study (FOS) | | | | | | | | | | | | | | | | | | Total |
|---|---|---|---|---|---|---|---|---|---|---|---|---|---|---|---|---|---|---|---|---|
| | | Art | Biology | Business | Chemistry | Computer science | Economics | Engineering | Environment | Geography | Geology | History | Mathematics | Medicine | Philosophy | Physics | Political science | Psychology | Sociology | |
| Falsification | n | 1 | 39 | 11 | 5 | 117 | 28 | 25 | 6 | 6 | 2 | 2 | 77 | 22 | 19 | 15 | 17 | 182 | 47 | 621 |
| | % | 100 | 81.25 | 91.67 | 100 | 89.31 | 93.33 | 100 | 85.71 | 100 | 66.67 | 100 | 93.9 | 95.65 | 90.48 | 78.95 | 94.44 | 95.29 | 92.16 | 92 |
| Induction | n | 0 | 4 | 1 | 0 | 13 | 2 | 1 | 0 | 0 | 0 | 0 | 4 | 2 | 1 | 3 | 1 | 6 | 2 | 40 |
| | % | 0 | 8.33 | 8.33 | 0 | 9.92 | 6.67 | 4 | 0 | 0 | 0 | 0 | 4.88 | 8.7 | 4.76 | 15.79 | 5.56 | 3.14 | 3.92 | 5.93 |
| Corroboration | n | 0 | 9 | 0 | 0 | 5 | 0 | 0 | 0 | 0 | 0 | 0 | 3 | 1 | 1 | 0 | 1 | 5 | 0 | 25 |
| | % | 0 | 18.75 | 0 | 0 | 3.82 | 0 | 0 | 0 | 0 | 0 | 0 | 3.66 | 4.35 | 4.76 | 0 | 5.56 | 2.62 | 0 | 3.7 |
| Demarcation | n | 0 | 3 | 0 | 0 | 4 | 1 | 0 | 1 | 0 | 1 | 0 | 2 | 0 | 0 | 1 | 0 | 5 | 3 | 21 |
| | % | 0 | 6.25 | 0 | 0 | 3.05 | 3.33 | 0 | 14.29 | 0 | 33.33 | 0 | 2.44 | 0 | 0 | 5.26 | 0 | 2.62 | 5.88 | 3.11 |
| Total | N | 1 | 55 | 12 | 5 | 139 | 31 | 26 | 7 | 6 | 3 | 2 | 86 | 25 | 21 | 19 | 19 | 198 | 52 | 707 |
| | % | 100 | 114.58 | 100 | 100 | 106.11 | 103.33 | 104 | 100 | 100 | 100 | 100 | 104.88 | 108.7 | 100 | 100 | 105.56 | 103.66 | 101.96 | 104.74 |
| Cases | N | 1 | 48 | 12 | 5 | 131 | 30 | 25 | 7 | 6 | 3 | 2 | 82 | 23 | 21 | 19 | 18 | 191 | 51 | 675 |

Notes. Results of chi-square tests for single concepts (rows): falsification: $\chi^2=23.52$, $p=0.533$; induction: $\chi^2=12.96$, $p=1$; corroboration: $\chi^2=37.56$, $p=0.011$; demarcation: $\chi^2=19.06$, $p=1$. Result of the overall chi-square test: $\chi^2=113.27$, $p=0.21$.

Table 4 shows the results of the CCA for *The logic of scientific discovery* (in German: *Logik der Forschung: Zur Erkenntnistheorie der modernen Naturwissenschaft*) and *conjectures and refutations: the growth of scientific knowledge* by Popper. "Falsification" is the concept which has been most frequently cited (92%). The overall chi-square test for the relationship between concept and FOS is not statistically significant (see the notes of Table 4). However, the partial results for "corroboration" are statistically significant. The results point out that the concept has been more frequently cited in biology than in other FOS.

Table 5. Citation concept analysis of *The logic of scientific discovery* (in German: *Logik der Forschung: Zur Erkenntnistheorie der modernen Naturwissenschaft*) and *Conjectures and refutations: the growth of scientific knowledge* by Popper: How frequently have the concepts (the concepts are decreasingly sorted by the column "Total") been cited in various time periods (publication years)?

| Concept | | Time period (publication years) | | | | |
| --- | --- | --- | --- | --- | --- | --- |
| | | <2000 | 2000-2005 | 2006-2010 | 2011-2018 | Total |
| Falsification | n | 101 | 118 | 209 | 193 | 621 |
| | % | 84.87 | 88.72 | 95 | 95.07 | 92 |
| Induction | n | 12 | 10 | 8 | 10 | 40 |
| | % | 10.08 | 7.52 | 3.64 | 4.93 | 5.93 |
| Corroboration | n | 7 | 10 | 6 | 2 | 25 |
| | % | 5.88 | 7.52 | 2.73 | 0.99 | 3.7 |
| Demarcation | n | 5 | 2 | 7 | 7 | 21 |
| | % | 4.2 | 1.5 | 3.18 | 3.45 | 3.11 |
| Total | N | 125 | 140 | 230 | 212 | 707 |
| | % | 105.04 | 105.26 | 104.55 | 104.43 | 104.74 |
| Cases | N | 119 | 133 | 220 | 203 | 675 |

Notes. Results of chi-square tests for single concepts (rows): falsification: $\chi^2$=15.45, *p*=0.006; induction: $\chi^2$=6.73, *p*=0.324; corroboration: $\chi^2$=11.81, *p*=0.032; demarcation: $\chi^2$=1.69, *p*=1. Result of the overall chi-square test: $\chi^2$=35.17, *p*=0.009.

With respect to Popper, Table 5 shows the relationship between the four concepts and various time periods. The overall chi-square test reveals statistically significant results: the concepts have been used with different frequency in the various time periods. The statistically significant overall result especially concerns the concepts "falsification" (increasing use over time) and "corroboration" (decreasing use over time).

Table 6. Importance of *The structure of scientific revolutions* by Kuhn and *The logic of scientific discovery* (in German: *Logik der Forschung: Zur Erkenntnistheorie der modernen Naturwissenschaft*) and *Conjectures and refutations: the growth of scientific knowledge* by Popper in different fields of study (the FOS are decreasingly sorted by absolute differences between the percentages; the book with the greater importance is tagged with K=Kuhn or P=Popper)

| FOS | Kuhn (1962) | | Popper (1934, 1959, 1962) | | Absolute difference between percentages |
|---|---|---|---|---|---|
| | $N$ | % | $n$ | % | |
| Sociology | 575 | 17.23 | 52 | 7.36 | 9.87 (K) |
| Mathematic | 132 | 3.95 | 86 | 12.16 | 8.21 (P) |
| Psychology | 706 | 21.15 | 198 | 28.01 | 6.86 (P) |
| Political science | 311 | 9.32 | 19 | 2.69 | 6.63 (K) |
| Economics | 343 | 10.28 | 31 | 4.38 | 5.89 (K) |
| Computer science | 480 | 14.38 | 139 | 19.66 | 5.28 (P) |
| Biology | 140 | 4.19 | 55 | 7.78 | 3.59 (P) |
| Engineering | 183 | 5.48 | 26 | 3.68 | 1.80 (K) |
| Philosophy | 62 | 1.86 | 21 | 2.97 | 1.11 (P) |
| Business | 92 | 2.76 | 12 | 1.70 | 1.06 (K) |
| Physics | 68 | 2.04 | 19 | 2.69 | 0.65 (P) |
| Geography | 47 | 1.41 | 6 | 0.85 | 0.56 (K) |
| Environment | 17 | 0.51 | 7 | 0.99 | 0.48 (P) |
| Medicine | 134 | 4.01 | 25 | 3.54 | 0.48 (K) |
| Chemistry | 13 | 0.39 | 5 | 0.71 | 0.32 (P) |
| Material science | 10 | 0.30 | 0 | 0.00 | 0.30 (K) |
| History | 1 | 0.03 | 2 | 0.28 | 0.25 (P) |
| Art | 12 | 0.36 | 1 | 0.14 | 0.22 (K) |
| Geology | 12 | 0.36 | 3 | 0.42 | 0.06 (P) |
| Total | 3338 | 100.00 | 707 | 100.00 | |

Note. Result of the overall chi-square test: $\chi^2$=225.72, $p$=0.000.

Table 6 shows the distribution of concept mentions of Kuhn and Popper in citances across FOSs (citations in citing papers with mentions of any of the concepts in the citation context data) – without considering the individual concepts. The data are from Table 2 and Table 4 (row: total *N*). We are interested in the meaning of both books in various FOSs. The results of the chi-square test points out that the books are differently used (statistically significantly) in the FOSs. The table is decreasingly sorted by the absolute differences between the percentages for both books in the table. The results reveal the largest difference for sociology: in general, Kuhn (1962) seems to be significantly more important (17.23%) as a conceptual base than Popper (1934, 1959, 1962) (7.36%). In mathematics, we observe the opposite (3.95% versus 12.16%).

Abbott (2016) analyzed citations to Kuhn (1962) with respect to the scientific disciplines of the citing papers. He identified three main audiences of Kuhn: (i) philosophy, history, and philosophy of science, (ii) social sciences and psychology, and (iii) applied fields. Our results confirm the high impact Kuhn's concepts have had in sociology and psychology (although our results show that Popper's concepts are more important in psychology). The high impact of Kuhn (1962) Abbott (2016) observed in history and philosophy might be less concept-specific according to our results. We also found high impact of Kuhn's concepts in applied fields, such as computer sciences, economics, and political sciences.

### 6.2   Second empirical part: uncertainty associated with concepts

Following an approach recently proposed by Small (2018) and Small et al. (2019), we measured uncertainty which might be associated with a certain concept. In the history of every scientific field, many concepts have been proposed. The literature dealing with these concepts provides information about the validity of the concepts: did the authors citing the concept confirm or question them? Using certain hedging words, we tried to catch uncertainty connected with the concepts investigated here in the citing literature. We expect that the conceptual approaches by Kuhn and Popper are differently associated with uncertainty. We also speculate that some concepts, but not others, are seen as certain knowledge. The results of the analyses considering various FOSs are shown in Table 7 for Kuhn, and Table 8 for Popper. In order to receive meaningful and interpretable results, only FOSs are considered with at least 300 concept citations in total. This meant for the CCA of Popper that FOS-specific analyses were not possible at all. The results in Table 7 reveal that around 14% of Kuhn's concept mentions are connected to uncertainty; for Popper, the percentage is around 12% (see Table 8). Thus, it seems that Kuhn's conceptual approach is slightly more frequently associated with uncertainty than Popper's conceptual approach (in all FOSs).

With around 22%, "crises" is the concept of Kuhn which is more frequently associated with uncertainty than all other concepts; with around 8%, "structure" is associated with the lowest rate of uncertainty. In the case of Popper, "corroboration" (16%) seems to be a more uncertain concept than "induction" (2.5%). For Kuhn, we also have FOS-specific results: in general, the most critical view on Kuhn can be observed in psychology (16.15%); the opposite is visible in economics (10.5%). "Paradigm" is most frequently seen as uncertain in psychology (16.5%), and less frequently in economics (11.31%). Computer science has the most critical view on "scientific revolution" (14.29%); for political science and sociology, the percentage is much lower with 6.25%.

In another study, Atanassova, Rey, and Bertin (2018) reported more uncertainty and hedging word occurrences in bio-medical disciplines than in physics. The corresponding FOSs received too few concept citations in our study to draw a meaningful comparison.

Table 7. Counts (*n* and *N*) and percentage of citances – referring to the concepts in *The structure of scientific revolutions* by Kuhn – reflecting uncertainty in selected fields of study (FOSs; only FOSs are considered with at least 300 concept citations in total; the column "Total" refers to all – not only the selected – FOSs; the concepts are decreasingly sorted by the column "*N*")

| Concept | Field of study (FOS) | | | | | | | | | | | |
|---|---|---|---|---|---|---|---|---|---|---|---|---|
| | Computer science | | Economics | | Political science | | Psychology | | Sociology | | Total | |
| | *n* | % | *n* | % | *n* | % | *n* | % | *n* | % | *N* | % |
| Paradigm | 308 | 14.29 | 221 | 11.31 | 210 | 14.76 | 503 | 16.50 | 403 | 12.66 | 2294 | 14.47 |
| Scientific revolution | 70 | 14.29 | 35 | 8.57 | 32 | 6.25 | 75 | 12.00 | 48 | 6.25 | 359 | 10.31 |
| Normal science | 44 | 11.36 | 39 | 15.38 | 27 | 3.70 | 36 | 19.44 | 48 | 14.58 | 254 | 12.99 |
| Structure | 24 | 8.33 | 12 | 0.00 | 19 | 5.26 | 33 | 9.09 | 26 | 3.85 | 160 | 8.13 |
| Anomaly | 14 | 28.57 | 21 | 4.76 | 11 | 36.36 | 30 | 16.67 | 17 | 23.53 | 113 | 18.58 |
| Incommensurability | 14 | 14.29 | 9 | 11.11 | 6 | 33.33 | 18 | 11.11 | 25 | 20.00 | 99 | 16.16 |
| Crisis | 6 | 16.67 | 6 | 0.00 | 6 | 0.00 | 11 | 45.45 | 8 | 25.00 | 59 | 22.03 |
| Total | 480 | 14.17 | 343 | 10.50 | 311 | 13.18 | 706 | 16.15 | 575 | 12.70 | 3338 | 13.93 |

Note. The table shows the (total) number of concept mentions in various FOS and – from that – the percentage of citances reflecting uncertainty. Based on the number of citances reflecting uncertainty (not shown), an overall chi-square test has been undertaken investigating the relationship between field-specific uncertainty and concept. The result of the test is $\chi^2=63.93$, $p=0.479$.

Table 8. Counts (*n*) and percentage of citances – referring to the concepts in *The logic of scientific discovery* (in German: *Logik der Forschung: Zur Erkenntnistheorie der modernen Naturwissenschaft*) and *Conjectures and refutations: the growth of scientific knowledge* by Popper – reflecting uncertainty (all fields of study, FOSs; the concepts are decreasingly sorted by the column "*n*")

| Concept | *n* | % |
|---|---|---|
| Falsifikation | 621 | 12.24 |
| Induction | 40 | 2.50 |
| Corroboration | 25 | 16.00 |
| Demarcation | 21 | 4.76 |
| Total | 707 | 11.60 |

Note. The table shows the (total) number of citances in various FOS and – from that – the percentage of citances reflecting uncertainty.

Table 9. Counts (*n* and *N*) and percentage of citances – referring to the concepts in *The structure of scientific revolutions* by Kuhn – reflecting uncertainty (the concepts are decreasingly sorted by the column "Total") in various time periods (aggregated publication years)

| Concept | | Time period (publication years) | | | | |
|---|---|---|---|---|---|---|
| | | <2000 | 2000-2005 | 2006-2010 | 2011-2018 | Total |
| Paradigm | *n* | 391 | 422 | 771 | 710 | 2294 |
| | % | 15.60 | 15.40 | 14.01 | 13.80 | 14.47 |
| Scientific revolution | *n* | 58 | 66 | 121 | 114 | 359 |
| | % | 17.24 | 9.09 | 9.09 | 8.77 | 10.31 |
| Normal science | *n* | 37 | 46 | 103 | 68 | 254 |
| | % | 13.51 | 10.87 | 13.59 | 13.24 | 12.99 |
| Structure | *n* | 23 | 34 | 54 | 49 | 160 |
| | % | 8.70 | 11.76 | 7.41 | 6.12 | 8.13 |
| Anomaly | *n* | 24 | 27 | 30 | 32 | 113 |
| | % | 25.00 | 25.93 | 13.33 | 12.50 | 18.58 |
| Incommensurability | *n* | 14 | 23 | 31 | 31 | 99 |
| | % | 0.00 | 30.43 | 16.13 | 12.90 | 16.16 |
| Crisis | *n* | 6 | 12 | 18 | 23 | 59 |
| | % | 16.67 | 16.67 | 16.67 | 30.43 | 22.03 |

| Total | *N* | 553 | 630 | 1128 | 1027 | 3338 |
|---|---|---|---|---|---|---|
| | % | 15.37 | 15.24 | 13.21 | 13.15 | 13.93 |

Note. The table shows the (total) number of concept mentions in four time periods and – from that – the percentage of citances reflecting uncertainty. Based on the number of citances reflecting uncertainty (not shown), an overall chi-square test has been undertaken investigating the relationship between time-specific uncertainty and concept. The result of the test is $\chi^2$=56.38, *p*=0.386.

In the context of the uncertainty analyses of the concepts by Kuhn and Popper, we also investigated developments over time. We are interested in whether the concepts are associated with an increasing or decreasing uncertainty. Since the case numbers for an analysis of the concepts introduced by Popper are too low, we refrain from the analysis on the basis of single concepts. The results for *The structure of scientific revolutions* by Kuhn in Table 9 show that the uncertainty decreases over time (from 15.37% before 2000 to 13.15% in 2011 to 2018). This decreasing trend is also visible for various concepts in the table (e.g., "paradigm" and "scientific revolution"). However, the reverse trend is also visible, e.g. for "crisis".

Table 10 shows the results for the development over time for Popper. With the exception of the time period before 2000, the results also reveal, similar to Kuhn, a decreasing rate of uncertainty.

Table 10. Counts (*n*) and percentage of citances – referring to *The logic of scientific discovery* (in German: "*Logik der Forschung: Zur Erkenntnistheorie der modernen Naturwissenschaft*") and *Conjectures and refutations: the growth of scientific knowledge* by Popper – reflecting uncertainty (increasingly sorted by the column "Time period", aggregated publication years).

| Time period | *n* | % |
|---|---|---|
| <2000 | 125 | 10.40 |
| 2000-2005 | 140 | 12.86 |
| 2006-2010 | 230 | 12.17 |
| 2011-2018 | 212 | 10.85 |
| Total | 707 | 11.60 |

Note. The table shows the (total) number of citances in four time periods and – from that – the percentage of citances reflecting uncertainty.

# 7   Discussion

The use of citation counts in research evaluation is based on the norm that "not citing relevant prior work constituted a violation" (Zuckerman, 2018). However, the focus on citation counts in quantitative research evaluation involves the danger of forgetting that a citation does not stand alone. It is embedded in the citing text which leads to different meanings of citations: "Authors may even cite articles that they consider as wrong, because they wish to correct their errors. In experimental disciplines, articles describing methods and apparatuses are also favoured. By contrast, big scientific discoveries, abstract new ideas, are rarely cited through the big original article, but rather through daughter publications inspired by the first" (Laloë & Mosseri, 2009, p. 28). The determination of different meanings is the objectives of citation context analyses. One goal of these studies is to develop methods allowing more detailed and informative citation analyses: which are, e.g., the papers in a specific field introducing important concepts which are frequently, intensively, and controversially discussed in a field? Quantitative (e.g., by using natural language processing techniques) and qualitative (e.g., classic close reading method) approaches can be performed for citation context analyses (Petrovich, 2018a).

In recent years, the full text of papers are increasingly available electronically which opens up the possibility of quantitatively investigating citation contexts in more detail. According to Lamers, van Eck, Waltman, and Hoos (2018) "this full text context of citations opens up many new research opportunities in citation analysis and may help reveal fundamental properties of and patterns in the process of knowledge accumulation in the sciences. Of particular interest … is to determine if systematic analysis of citation context can help shed light on what role previous literature plays when cited in new publications, how new authors use past literature to further their own arguments, and whether we can disentangle disciplinary modes of knowledge accumulation from more general archetypes of contributions made to the scientific landscape". Following the notion by Bu et al. (2019) of a multi-dimensional perspective on citation impact, we introduced in this study the CCA method for determining the citation impact certain published concepts have had on the scientific community. The CCA allows citation context analyses which fit perfectly into the following statement by Petrovich (2018a) that "citation context analysis seems particularly suitable to clarify the fine-grained structure of the knowledge accumulation process" (p. 1127). CCA can be used to determine the importance of scientific concepts in the so-called Third World (Popper, 1972). This is the world of "objective knowledge" which can be separated from "the First World (the physical world) and the Second World (the mental world)" (Petrovich, 2018b, p. 11.10).

It is worth reflecting on which of the concepts in Kuhn's and Popper's landmark books have left an impact, and on which fields of study. It is not surprising that Kuhn's "paradigm" concept has had a significant impact. What is surprising is that it has had such a disproportionately larger impact than Kuhn's other concepts. The paradigm concept accounts for over 80% of the citations to Kuhn's work. And its impact is resilient across all FOSs and over time. It is also not surprising that the second most popular concept derived from Kuhn's book is "scientific revolution". Kuhn argued that revolutions are an integral part of the development of a scientific field. Even though his concern was with the natural sciences, many scholars in the social sciences looked at their own fields of study in an effort to determine the extent to which Kuhn's model of scientific development describes their own field (see Wray, 2017).

Further investigation is needed in order to explain our finding that the concept of "normal science" has had a greater impact in mathematics than physics. One conjecture is that conceptual innovations in mathematics augment the existing knowledge, rather than replace what was regarded as secure knowledge before. For example, the discovery or development of non-Euclidean geometries did not render Euclidean geometry obsolete. In this way, all innovations in mathematics are continuous with a single normal scientific research tradition. But developments in physics are not like this. Newton's physical theory led to the rejection of the contact physics associated with Galileo and Descartes. About two centuries later, Albert Einstein's theory of general relativity led to a generalization of Newton's gravitational theory. And the photon theory of light had a similar effect on the theories of light accepted by earlier generations of physicists. Normal scientific research traditions are interrupted in physics, as new theories replace older theories. Indeed, it is worth remembering that Kuhn's cyclical theory of scientific change was explicitly designed with the natural sciences in mind. He did not purport to be describing the dynamics of conceptual change in either the formal sciences, like mathematics, or the social sciences.

Similarly, with respect to Popper, it is not surprising that "falsification" would be the most used concept derived from his books. Falsification, after all, is the cornerstone of Popper's critical rationalism. According to Popper (1934, 1959, 1962), the only way scientists can advance our scientific knowledge is by (i) attempting to falsify hypotheses, and (ii) rejecting those that are falsified. Falsifiability is also, according to Popper (1934, 1959, 1962), what distinguishes a science from a pseudo-science. But, again, it is surprising that citations to falsification account for over 90% of the citations to his books. Further investigation would be needed in order to explain why "corroboration" is used more in biology than in other disciplines.

Our finding, that Kuhn's book has a greater comparative impact than Popper's books in sociology, is not surprising for two reasons. First, Kuhn (1962) develops a theory of science and scientific change that gives special attention to the changing social dynamics in scientific fields (see Wray, 2011). The breakdown of the consensus in a scientific field that characterizes a pending scientific revolution or paradigm change is explicitly described in social terms. The field is described as being in crisis as the consensus breaks down. Even in Kuhn's discussion of scientific revolutions, he explicitly draws a comparison between scientific revolutions and political revolutions. It is not surprising that this sort of analysis would appeal to sociologists. Popper, on the other hand, focuses narrowly on analyzing the logic of science, in keeping with the positivist tradition in which he grew up. Second, Popper had expressed disdain for the social sciences (see Wray, 2017). Kuhn, on the other hand, avoided saying much about the social sciences.

Our finding that Popper's books had a greater impact than Kuhn's book in mathematics is also not surprising. Kuhn, after all, was only concerned with the empirical natural science, not formal sciences, like mathematics. And Popper explicitly discusses probability theory in his book, thus making it relevant to at least some mathematicians, specifically those working in probability theory.

The fact that the level of uncertainty associated with the key concepts in Kuhn's book decreases as we approach the present is not especially surprising. It could easily be a function of the fact that people now seem to have settled on how the terms should be interpreted or applied. Similar remarks explain the trends with respect the uncertainty associated with the use of the key terms in Popper's books. These books have existed long enough to have achieved canonical status in philosophy of science.

Our study has some limitations. The most severe limitation is probably that the MA database does not contain citation context information for all citing papers. For Popper, 14.4% of the citing papers have citation context information available. In the case of Kuhn, 12.2% of the citing papers have citation context information available. A less severe limitation is the selection of concepts and search terms for the concepts and the uncertainty detection. Here, also the wording and length of the citation context determines our ability to properly assign citation contexts to concepts and determine their uncertainty. Finally, the least severe limitation is the fact that some publications are not assigned to FOSs in MA. As 99.7% of papers citing Popper's books and 99.5% of papers citing Kuhn's book were assigned to a level 0 FOS, this limitation seems negligible.

FOSs in MA are assigned algorithmically on the paper-basis. The quality of the FOS assignment is unclear. Algorithmic FOS assignments may or may not be accurate. The accuracy of an algorithm based on direct citation relations has been questioned (Haunschild, Marx, French, & Bornmann, 2018; Haunschild, Schier, Marx, & Bornman, 2018). A case study on computer science publications reported promising results regarding the MA FOSs (Scheidsteger, Haunschild, Hug, & Bornmann, 2018). However, a large-scale comparison is still missing. Despite these limitations, we are confident that CCA is a very interesting method. The books by Kuhn and Popper were used to show how the CCA method can be applied. If the missing citation context information is distributed randomly (without bias towards concepts, uncertainty words, publication years, and FOSs) our results should remain valid despite the mentioned limitations.

## Acknowledgements

The bibliometric data used in this paper are from a locally maintained database at the Max Planck Institute for Solid State Research derived from the Microsoft Academic database. We would like to thank Henry Small for discussing the use of hedging words for measuring uncertainty. He also supports our study by providing us with his initial hedging words for measuring uncertainty. KBW's research is supported by a grant from Aarhus Universitets Forskningsfond – Starting Grant (AUFF): AUFF-E-2017-FLS-7-3.